# Margination of artificially stiffened red blood cells

Revaz D. Chachanidze,[1,2] Othmane Aouane,[3] Jens Harting,[3,4] Christian Wagner,[5,6,*] and Marc Leonetti[2,7,8,†]

[1]*Experimental Physics, Saarland University, 66123 Saarbrücken, Germany*
[2]*Aix Marseille Univ, CNRS, Centrale Marseille, IRPHE, Marseille, France*
[3]*Helmholtz Institute Erlangen-Nürnberg for Renewable Energy, Forschungszentrum Jülich, Cauerstraße 1, 91058 Erlangen, Germany*
[4]*Department of Chemical and Biological Engineering and Department of Physics, Friedrich-Alexander-Universität Erlangen-Nürnberg, Cauerstraße 1, 91058 Erlangen, Germany*
[5]*Experimental Physics, Saarland University, 66123 Saarbrucken, Germany*
[6]*Department of Physics and Materials Science, University of Luxembourg, L-1511 Luxembourg, Luxembourg*
[7]*Aix Marseille Univ, CNRS, Centrale Marseille, M2P2, Marseille, France*
[8]*Aix Marseille Univ, CNRS, CINAM, Turing Centre for Living Systems, Marseille, France*

(Dated: August 28, 2024)

Margination, a fundamental process in which leukocytes migrate from the flowing blood to the vessel wall, is well-documented in physiology. However, it is still an open question on how the differences in cell size and stiffness of white and red cells contribute to this phenomenon. To investigate the specific influence of cell stiffness, we conduct experimental and numerical studies on the segregation of a binary mixture of artificially stiffened red blood cells within a suspension of healthy cells. The resulting distribution of stiffened cells within the channel is found to depend on the channel geometry, as demonstrated with slit, rectangular, and cylindrical cross-sections. Notably, an unexpected central peak in the distribution of stiffened RBCs, accompanied by fourfold peaks at the corners, emerges in agreement with simulations. Our results unveil a non-monotonic variation in segregation/margination concerning hematocrit and flow rate, challenging the prevailing belief that higher flow rates lead to enhanced margination.

*Introduction.–* In physiological blood flow, leukocytes partake in a phenomenon known as "margination"[1–5]. This process plays a pivotal role in the immune response by enabling the interaction between white blood cells and the vessel wall, facilitating their subsequent migration to sites of inflammation [6, 7]. Blood primarily comprises red blood cells (RBCs) constituting 45% of its volume, suspended in a predominantly Newtonian buffered protein solution called plasma. While other cell types, such as white blood cells, make up a mere fraction of the total volume, margination ensures that they tend to flow closer to the vessel wall. This phenomenon is believed to be driven by differences in size and stiffness among the various cell types. Moreover, margination is not confined to white blood cells; it also manifests in platelets and in pathological conditions like malaria and sickle cell disease, where rigidified RBCs exhibit margination [8–10]. From a physical perspective, it raises the question of how a binary mixture of flexible objects of different stiffnesses self-organizes in the non-equilibrium situation of a pressure-driven flow.

Ordering of stiff particles in a highly dilute duct flow at low but finite Reynolds number ($Re \gtrsim 1$) is described by the Segré-Silberberg effect [11]. The salient features are the shear-induced migration force and the inertial lift force experienced by rigid particles in bounded flow [12–16]. Moreover, efficient multi-particle separation, still at low concentrations, is achieved at the micrometric scale by coupling the inertial lift force to Dean drag forces in numerous devices [17]. In a square channel, a rigid sphere is localized at four-point attractors close to the centers of each facet/side [18]. Considering suspensions in an inertial regime, rigid particles can also accumulate in these fourfold symmetric points, leading to pattern formation [19].

In the limit of vanishing Reynolds number ($Re \ll 1$), symmetry arguments of the Stokes equation preclude any lateral migration across streamlines of solid particles. This mathematical statement breaks down for a soft particle, which, while deforming, breaks the fore-aft symmetry, leading to a viscous lift force that pushes the particle away from the wall. The softer it is, the greater the viscous lift force, a phenomenon highlighted by studying the dynamics of vesicles, microcapsules, and red blood cells [20–24]. In a monodisperse suspension of such soft core-shell particles, short- and long-range hydrodynamic interactions in a shear flow induce hydrodynamic diffusion, spreading any density heterogeneity in the bulk [25, 26]. However, these physical effects lead to the creation of a particle-free layer at the periphery of the flow of a monodisperse suspension of soft core-shell particles: the so-called "Cell-Free Layer" (CFL) in blood vessels. This phenomenon is well understood experimentally in blood flow [27–29] or suspensions of physiological and stiffened RBCs [30], theoretically by coarse-grained models [31, 32], and numerically in biomimetic suspensions made of vesicles, microcapsules, and models of RBCs, thanks to the growing interest and intense activity of the High-Performance Computing (HPC) community in the past decade [30, 33–37]. For the case of a binary dense suspension, i.e., white and red blood cells, the situation is less clear, and experiments are scarce and controversial. Aggregation of RBCs promotes margination [38], while recently, it was claimed that the viscoelastic properties of plasma (suspending fluid) are at the origin of margination [39]. From a theoretical perspective, it was found that shear-induced pair collisions should be the most dominant cause for margination [40, 41]. Several numerical simulations of a mixture of two kinds of soft core-shell particles in bounded Poiseuille flow have demonstrated segregation of both species [42–47], but generally, they differ in shape, size, and stiffness simultaneously.

---
[*] christian.wagner@uni-saarland.de
[†] marc.leonetti@univ-amu.fr

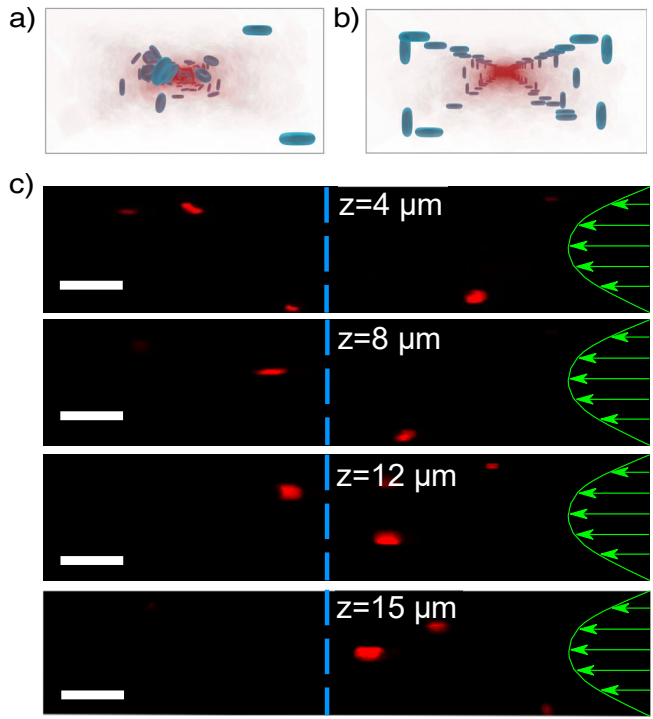

Figure 1. (Color online) (a) and (b): numerical simulations. (c): experiments. All conducted in a channel with a height of $h = 30$, $\mu$m and $w = 60 \mu$m. (a) Initially, soft (light red) and stiff (blue) particles with the typical discoidal reference shape of RBCs are distributed randomly. (b) At steady-state, the stiff cells are positioned in the four corners of the channel. (c) Confocal images show fluorescently labeled stiffened RBCs (red) at different z-focal planes. Scale bar = 5 $\mu$m.

Both particle size and shape play important roles in determining margination behavior. Non-spherical particles, particularly those with a high aspect ratio, tend to marginate more efficiently than spherical particles [48, 49]. This is due to the role of particle rotation in the margination process, which facilitates interactions with the vessel wall [50]. For example, oblate and rod-shaped particles show enhanced margination compared to spherical particles, especially in the presence of RBCs, which promote their migration toward the vessel wall [49]. Particle size also plays a crucial role, with several studies suggesting an optimal size for margination. Larger particles in the micrometer range tend to marginate more effectively [4], while the results for smaller particles have been mixed. Spherical particles larger than 500 nm exhibit pronounced margination, while those smaller than 200 nm are likely to remain trapped between RBCs in the core of the blood flow, away from vessel walls [51]. However, some studies have shown favorable margination behavior for nano-sized particles (< 100 nm) due to the influence of Brownian motion [48, 52]. Additionally, particles around one micron in diameter have been found to exhibit the highest likelihood of forming ligand-receptor bonds with vessel walls [50]. The impact of particle stiffness on margination is still an area of active research, with conflicting results reported in the literature. Some studies have shown that RBC elasticity does not significantly affect the margination of white blood cells (WBCs) [53]. In contrast, others have demonstrated that collisions between stiff and elastic particles enhance margination [54, 55]. Experimental work using stiffened cells in animal models has also shown that increased particle stiffness leads to a higher degree of margination [56–58]. However, the isolated effect of particle stiffness remains an open question, as most studies involve interactions with other factors such as size and shape.

Here, we address the problem of margination combining in vitro and in silico investigations (Fig.1). We quantify blood flow consisting of two populations of red blood cells – healthy and rigidified with the cross-linking agent (glutaraldehyde)[59] – in microfluidic channels for different flow rates, hematocrits, and vessel geometries. RBCs are soft core-shell particles with a discoidal shape. They are approximately $8\mu$m in diameter and $3\mu$m thick. Such an experimental model allows us to examine margination caused exclusively by the rigidity contrast between two subpopulations of cells. We use confocal microscopy to reconstruct 3D distributions of labeled cells across the section [5]. We identify similar observations *in vitro* and in silico: similar patterns of segregation and entrance lengths, as well as the presence of an unexpected maximum of segregation according to the flow rate and the volume fraction ($H_t$) of physiological RBCs.

*Materials and Methods.–* Whole blood is taken from healthy volunteers via venipuncture and collected into vacutainer tubes containing EDTA. First, blood is centrifuged (Hermle Labortechnik GmbH Z36HK) at 1400g for 10 minutes to separate RBCs from plasma and other blood components. After that, the RBC pellet is thoroughly washed 3 times in physiological buffer solution (PBS, Gibco Life Technologies), acquired as a stock solution and not pills, containing 10 mg/ml of BSA and 5.5mM of glucose (Sigma-Aldrich). Washed cells are finally resuspended and adjusted to the required hematocrit in an isodense iodixanol-based solution (OptiPrep™, Sigma-Aldrich). This suspending medium prevents RBCs from sedimentation and contains 35% v/v concentration of OptiPrep™ in PBS. A rheological characterization with a rotational rheometer (MCR702, Anton Paar, Austria) showed no relevant differences between the samples in PBS and OptiPrep™/PBS mixtures [60]. The necessary RBC concentration is achieved by carefully pipetting a required volume of the washed RBC pellet. The hematocrit level of a blood sample is verified with a micro-hematocrit centrifuge. Then, a small portion of RBCs (<1%) is rigidified with glutaraldehyde (Sigma-Aldrich, Saint Louis, USA) for 1 hour, thoroughly washed, and then fluorescently labeled with CellMask™ Deep Red Plasma membrane stain (Molecular Probes Life Technologies). Labeled stiff cells are added to the sample to create a binary suspension with two populations of cells with different rigidities.

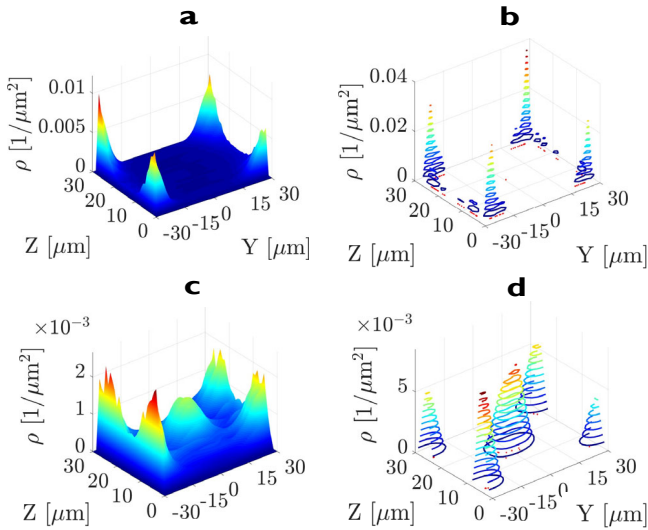

Figure 2. (Color online) The distribution of stiff cells at steady state. (a) Experimental results at $H_t$ = 20% hematocrit, the mean velocity of the flow is $v_m \approx 900\mu$m/s, and $Ca$ = 0.24. (b) Numerical results at $H_t$ = 20%, $v_m \approx 1300\mu$m/s, and $Ca$ = 0.2 for the soft RBCs and $Ca$ = 0.002 for the stiffened RBCs. (c) Experimental results at $H_t$ = 40% and $v_m \approx 2850\mu$m/s and $Ca$ = 0.76. (d) Numerical results at $H_t$ = 40%, $v_m \approx 500\mu$m/s, and $Ca$ = 0.2 for the soft RBCs and $Ca$ = 0.002 for the stiffened RBCs.

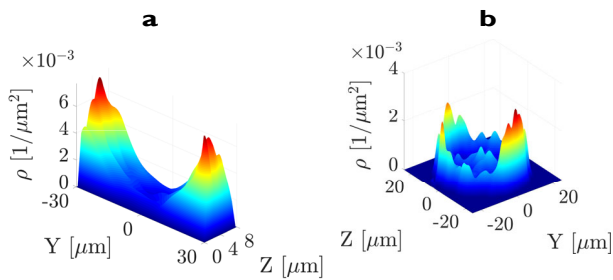

Figure 3. (Color online) The distribution of stiff cells at steady state. (a) Margination in a slit geometry obtained in an 8$\mu$m high and 60$\mu$m wide channel at $H_t$ = 20% hematocrit, a mean velocity of the flow $v_m \approx 900\mu$m/s, and a capillary number $Ca \approx 0.24$. (b) Segregation in a glass microcapillary of the inner diameter of 50$\mu$m at $H_t$ = 20% hematocrit, a mean velocity of the flow $v_m \approx 1000\mu$m/s and a capillary number $Ca \approx 0.32$.

AFM measurements showed that the fixed cells have a Young modulus that was at least three orders of magnitude larger than that of the healthy cells, i.e., they can be considered as completely rigid [61]. Fixed cells are kept as a stock solution at 4° C for an indefinite period in PBS containing 20 mg/ml of BSA to passivate any residuals of glutaraldehyde. Blood collection and realization of the experiment take place on the same day.

PDMS channels are constructed using a standard soft lithography technique. Confocal measurements are performed with a spinning disk-based confocal head (CSU-W1, Yokogawa Electric Corporation) coupled with an inverted microscope (Nikon Eclipse Ti). A solid-state laser (with a wavelength of 647nm, Nikon LU-NV Laser Unit) is used as a light source for irradiating fluorescently labeled cells. Image sequences are acquired with a digital camera (Orca-Flash 4.0, Hamamatsu Photonics, Hamamatsu City, Japan).

Numerical simulations are performed using our in-house hybrid lattice Boltzmann/immersed boundary 3D code. A finite element method is incorporated to calculate the elastic force the red blood cells exert on the fluid. Our code has consistently predicted the complex behavior of soft and rigid particles in simple and complex geometries in the past [30, 41, 62–64]. A detailed description of the algorithm is provided in our previous works [63, 65], and the numerical setup is described in the supplementary material section [61]. For this paper, we simulated up to 18000 red blood cells to achieve the largest volume fraction (40%). Each cell is discretized with a triangular mesh with 2000 faces and 1002 nodes. Although our code is parallelized, some of the simulations corresponding to low velocities and large volume fractions require run times of several weeks on a little more than 4000 CPU cores.

Image processing of the experimental photographs and the evaluation of the numerical data are performed using the MATLAB™(MathWorks, MA, USA) computing environment. Apart from the velocity calculation of labeled cells, this custom-developed MATLAB software is used to evaluate the coordinates of the moving cells and their velocity-corrected probability density as a function of the distance from the wall. To characterize the margination strength (wall concentration in percent), the ratio of the integrated probability density from the wall to 8$\mu$m into the bulk is compared to the integrated probability density in the inner part of the channel.

The hydrodynmic behavior of a red blood cell with a radius of $R$ and a shear elastic modulus of $G_s$ is influenced by two dimensionless parameters. The first parameter is the Reynolds number at the scale of the particle, which measures the ratio of inertial forces to viscous forces and is defined as $Re_p \equiv Re(\frac{R}{D_h})^2$, with $D_h$ being the hydraulic diameter. In our experiments, $R_p$ does not exceed 0.007, while it is fixed to 0.5 in numerical simulations. The second important dimensionless parameter is the capillary number. It describes the ratio of external shear stress to the deformability of the red blood cell. It is defined as $Ca \equiv \rho v R \dot{\gamma}_w / G_s$, with $\dot{\gamma}_w$ representing the wall shear rate.

While *Ca* is fixed to approximately 0.2 for physiological RBCs in the numerical simulations, its value varies in the experiments between 0.05 and 1.2.

*Results.*–First, we consider a hematocrit of $H_t = 20\%$ and rectangular microchannels in two limiting cases, i.e., with aspect ratios such as $h/w << 1$ and $h/w \approx 1$ where $h$ and $w$ are the height and the width, respectively. In rectangular channels of $h = 30\mu m$ and $w = 60\mu m$, the distribution of stiff RBCs is evaluated employing confocal microscopy (see Fig. 1(c)). Cells passing through the section of a channel are tracked in each focal plane, and the final probability density function is shown in Fig. 2(a). It is noteworthy that segregation appears at the four corners where the shear rate within the near to the wall regions is minimal. This statement is assessed by considering other geometries. In rectangular microchannels with a slit-like geometry (aspect ratio $h/w = 8\mu m/60\mu m \approx 0.13$), a quasi mono-layer of particles flows along the slit. Again, we observe a significant level of cell segregation. Stiff particles accumulate near the side walls of the channel where the shear rate is highest (Fig. 3a). Finally, we consider a circular cross-section. A glass microcapillary of $50\mu m$ is set vertically along the optical path of the objective and is plunged into a transparent cuvette filled with a fluid of higher density to prevent any growing deposit of RBCs. This approach allows us to visualize flowing cells at a capillary's exit and directly estimate their spanwise distribution [61]. Stiff particles are mainly allocated along the capillary wall (Fig. 3(b) where the shear rate is at its maximum in contradiction with the experimental results of [39]. However, we also observe a small accumulation at the center, where the shear rate is minimal. These results can seem puzzling if the shear rate, a scalar quantity, is considered. Indeed, such a peak can also be found if we vary the flow rate and the hematocrit in a rectangular microchannel with $h/w = 0.5$. When the hematocrit is increased from 20% to 40%, again, the four previously evoked peaks of stiff particles appear at the four corners, and at the same time, the central peak appears where the shear rate vanishes (Fig.2(c)). This leads to a more complex view of margination.

In order to verify that only the differences in stiffness of the cells cause the margination, we perform numerical simulations of a binary suspension of soft and stiff microcapsules that flow in the rectangular channel with the same dimensions as in experiments, i.e., the one of aspect ratio $30/60 = 0.5$. As shown in Fig. 2(b and d), our numerical simulations confirm the qualitative picture with stiff particles gathering at the four corners. And very comparable to the experiments, the simulations predict as well a peak in the center of the flow when the hematocrit is increased to 40%.

To gain insight into the dynamics of segregation, we perform experiments to evaluate the entrance length, i.e., the distance required to establish a fully developed distribution of segregated stiff cells. For this purpose, we evaluate their distribution at different positions from the entrance of the large aspect ratio microchannel to a distance of 5cm for three different hematocrits. The applied pressure is adjusted to obtain the same flow rate ($U \approx 900\mu m/s$) in all three cases. The data are well fitted with a simple exponential function where the entrance length is the characteristic length, as shown in Fig. 4. The entrance length in the experiments, i.e., the distance to reach the saturation of the segregation pattern, is equal to 2.71, 0.67, and 0.55 cm for $H_t = 10$, 20, and 40%, respectively, i.e., typically two orders of magnitude larger than the width of the microchannel. These lengths are comparable to the ones that are needed to form the cell-free layer [30]. The saturation level in the experiments depends non-monotonically on the hematocrit $H_t$. This can be observed even better if we focus on the dependency of the margination on the velocity, as shown in Fig. 5. We find that segregation is most expressed at a mean flow velocity of $v_m \approx 1000~\mu m/s$. With further increase of the velocity, the degree of segregation is attenuated for all studied hematocrits. Cell segregation is obviously more expressed at 20% concentration of deformable cells. This is also well reproduced by our numerical simulations. Unfortunately, reaching full convergence of the simulations corresponding to low velocities and large volume fractions was not possible due to limited available computational resources. As such, we carried out the simulations at a higher speed of 5mm/s. We find a higher amount of cell segregation at a steady state in the simulations and assume that the main reason for this discrepancy is the fact that the cell-free layer (CFL) in simulations is literally free from any healthy and thus flexible cells. In contrast, in the experiments, we observe (in classical bright field mode) under the microscope, both more "abnormal" healthy cells, such as spherocytes, echinocytes, or trilobes in the CFL-region [66]. In other words, in the experiments, there is a certain contamination of rigid cells in the suspension of flexible cells, which might lead to a certain reduction of the absolute amount of margination. Despite this discrepancy, the numerical model captures the dependency of the degree of segregation on the concentration of the stiffened cells qualitatively well (inset of Fig. 5).

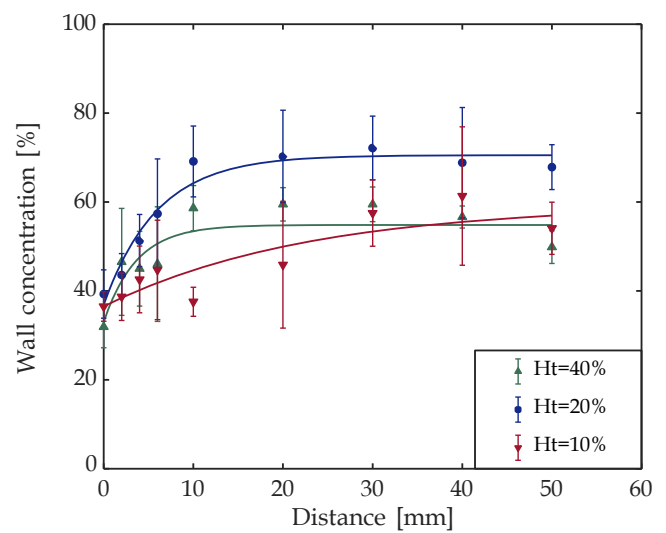

Figure 4. (Color online) Experimental results of the evolution of segregation (near wall concentration of stiffened cells) along the flow in the microchannel at different hematocrits. The mean flow velocity of flowing cells $v_m \approx 900\mu m/s$, corresponding to a capillary number $Ca \approx 0.24$. Error bars are 95% confidence interval of the mean calculated using the sample rate of distinct experiments as a weighting coefficient.

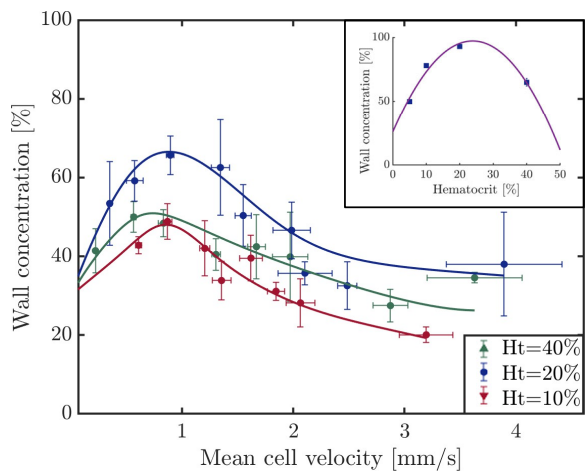

Figure 5. (Color online) Experimental results of the maximum segregation according to flow rate and hematocrit. Symbols are as in fig 4. Effect of varying velocity on cross-stream particle migration in experiments. Solid lines are only guides for the eye. Error bars are 95% confidence interval of the mean calculated using the sample rate of distinct experiments as a weighting coefficient. In the main figure, the capillary number $Ca$ varies approximately between 0.05 to 1.2. Inlet: Numerical results for the maximum segregation for different hematocrit at a centerline bare fluid velocity of $\bar{v} \approx 5000 \mu m/s$ and a capillary number $Ca = 0.2$ for the soft RBCs and $Ca = 0.002$ for the stiffened RBCs
.

*Summary.–* The segregation of rigid cells is studied experimentally and numerically in a wide range of flow velocities. Notably, our numerical simulations are in full qualitative agreement with our experimental findings. However, simulations of more than 10.000 cells are extremely costly in terms of computing resources, and a steady state can only be reached for a few runs at large velocities.

Our findings reveal an unexpected non-monotonic behavior in particle segregation, a phenomenon previously unobserved due to the constrained range of flow rates in earlier studies [4, 42, 43, 45, 67–70]. Rigid cells concentrate at the four corners of a rectangular channel and in the center of the flow, at least at higher hematocrit. Interestingly, these initially centrally located rigid cells seem to become "trapped" and resist lateral migration at elevated hematocrit values. A similar observation was done for the case of small spherical particles in a suspension of RBCs [5]. The lateral migration of rigid particles is found to depend non-monotonously on the volumetric fraction of the dominant population of deformable particles. The lateral migration of rigid particles is intricately linked to the volumetric fraction of deformable particles and exhibits non-monotonic behavior, a trend reminiscent of previous numerical [43] and experimental [38] studies. A mechanistic model for margination proposed in Ref.[71] underscores the role of two competing factors: wall-induced hydrodynamic migration for deformable particles and particle pair collisions, both homogeneous and heterogeneous. Collisions in this context does not necessarily mean that the cells get into direct contact, but they interact by their surrounding hydrodynamic stress field which is different for soft and for solid objects While this model sheds light on the migration of stiff particles toward the walls, their accumulation in the center and the non-monotonic behavior as a function of velocity or hematocrit remains a puzzle[41]. Furthermore, our results show that a two-dimensional shear gradient has a stronger effect on the margination than a one-dimensional one, even if the absolute value of the latter might be bigger.

In some experimental investigations, a link has been drawn between margination and rouleaux formation, the reversible aggregation of red blood cells induced by plasma proteins at low shear rates [38, 72]. However, in our case, rouleaux formation is conspicuously absent, and consequently, aggregation does not play a pivotal role in particle segregation within our experiments. This underscores the complexity of the margination phenomenon and the various factors at play.

Lastly, it's worth noting that both our simulations and experiments have highlighted that margination necessitates travel distances on the order of centimeters to achieve a steady state, a finding consistent with previous work [73]. These distances are considerably longer than the average spacing between bifurcations in in-vivo capillaries, arterioles, and venules. This underscores the need for future studies to incorporate more intricate flow geometries for a realistic depiction of margination. Of course, the behavior of white cells and platelets might also differ severely from the behavior of stiffened red blood cells, and a detailed study on their specific margination behavior should be performed in future investigations.

# I. ETHICAL STATEMENT

## I. SUPPLEMENTARY MATERIAL

### 1. *AFM measurements*

In order to quantify the rigidity contrast between two subpopulations of RBCs in imagination experiments, we employed AFM. The effective Young's modulus of cells was measured through the recording of force–distance curves. AFM measurements were performed in a liquid environment (PBS) using a JPK Nanowizard 3 setup coupled to an optical microscope. We investigated the rigidity of RBCs rigidified with a cross-linking agent (glutaraldehyde) [1]. Cells were fixed (rigidified) with a range of glutaraldehyde concentrations in order to quantify the dependence of cells' rigidity on cross-linking agents. A variety of cantilevers with various nominal spring constants, as well as different indentation forces, were tested in order to find proper measurement conditions for each glutaraldehyde concentration. Prior to measurements, cells were immobilized on a substrate with adhesive protein (CellTak). Force mapping was performed for 3-5 cells of each population on a grid of 32× 32 points, corresponding to 10× 10 $\mu$m map. Force-distance curves were acquired at an indentation rate of 5 $\mu$m/s. Curves were analyzed according to the Hertz model implemented in JPK software. Poisson ratio was set to 0.5. The force-distance mapping of healthy RBCs was performed with the same experimental protocol as rigid cells. It is possible that the immobilization of healthy RBCs on a substrate with CellTak could result in residual stress on the cell membrane. Nevertheless, within the limits of qualitative comparison of fixed and healthy cells, it is a justified assumption to consider this effect negligible. From these results, we can conclude that RBCs fixed in 0.1-1% of glutaraldehyde can be regarded as solid objects compared to healthy RBCs.

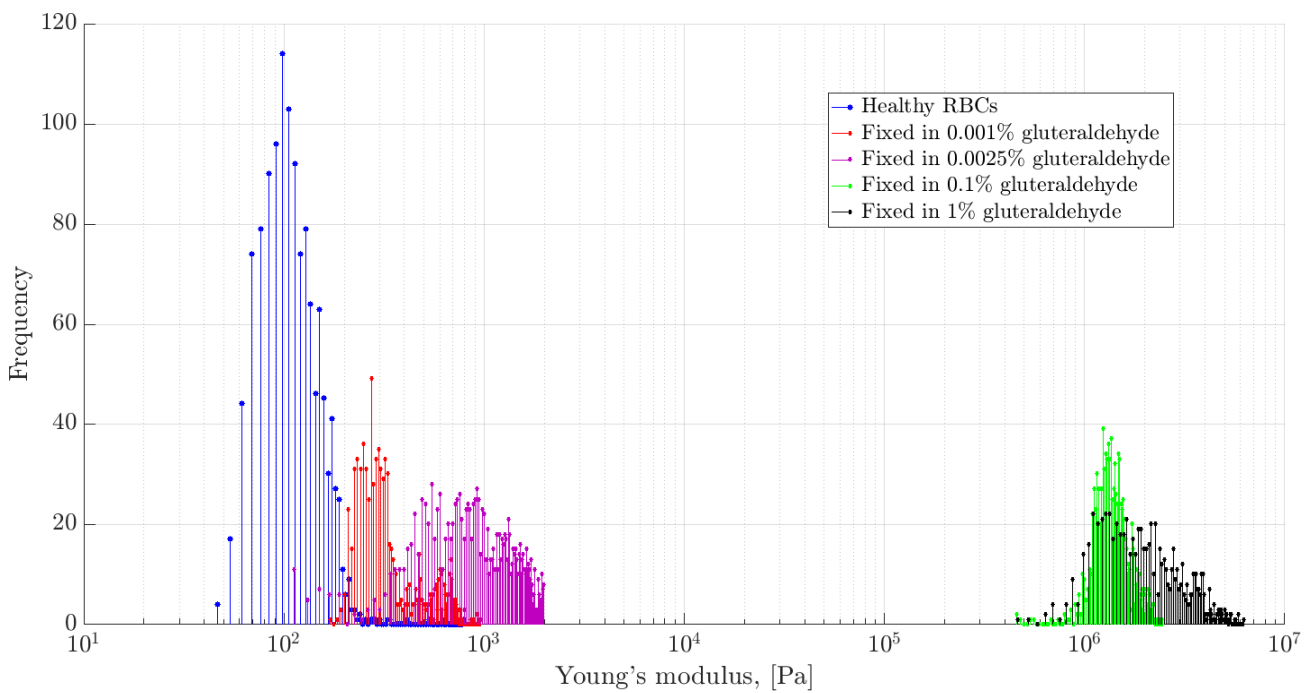

Figure 1. Results of AFM measurements demonstrate considerable contrast in rigidity between healthy RBCs and cells fixed in 0.1-1% of glutaraldehyde [2].

### 2. *Rheological characterization of the samples*

We have performed several experiments to determine the viscosity of our blood solution with ANTON Paar MCR 702 rheometer [2]. For these measurements we were using counter-rotating cylinders (the CC20 Taylor-Couette geometry) in order to minimise the sedimentation effect. The data were acquired for six hematocrit values between 5-50%, 4-5 blood samples each value. As a frame of reference, additional viscosity data for the whole blood were included to demonstrate the shear-thinning behaviour of unwashed suspension of RBCs. In this case the formation of aggregates (rouleaux) due to the presence of plasma proteins leads to the pronounced shear-thinning. In contrast to it, viscosity measurements of a suspension of rigidified RBCs with a volumetric fraction of 5% show an indisputable Newtonian behaviour. The samples were pre-sheared for one minute at ˙ = 100 s 1 before acquiring data. The viscosity was measured in the range of shear rate values between 0.1 - 500 s 1 at the grid of 21 points on logarithmic scale. The experimental data was acquired while gradually increasing shear-rate.

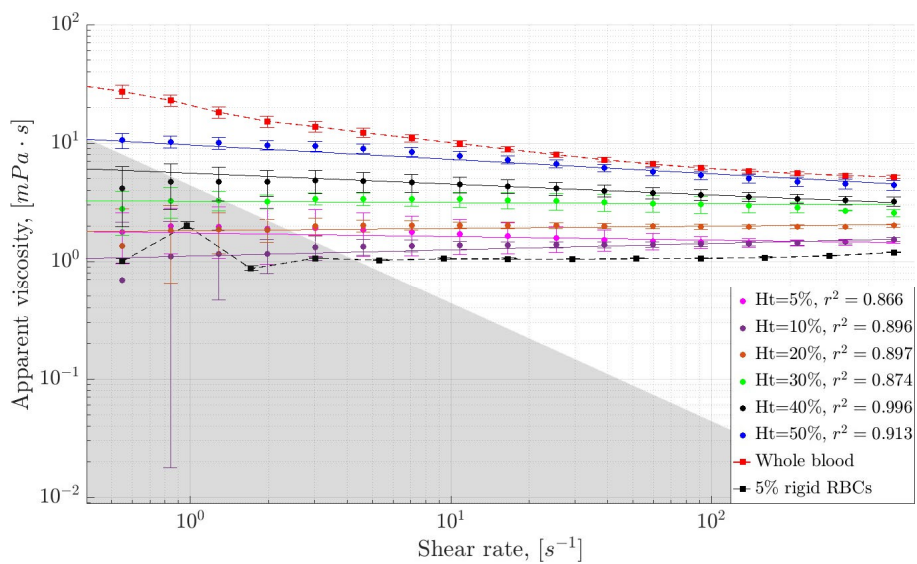

Figure 2. Rheological measurements of RBCs suspended in PBS. The grey area is the low-torque resolution limit of the ANTON Paar MCR 702 rheometer.

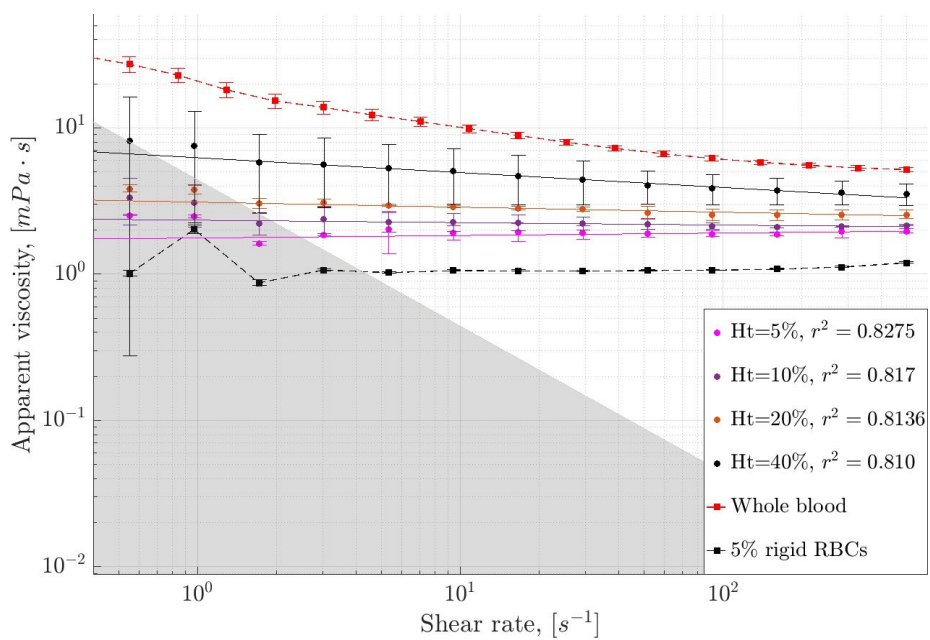

Figure 3. Rheological measurements of RBCs suspended in a media, composed of 65% PBS and 35% OptiPrep™. The grey area is the low-torque resolution limit of ANTON Paar MCR 702 rheometer.

## 3. *The lattice Boltzmann – Immersed Boundary method for blood flow*

The fluid flow is simulated using the lattice Boltzmann method (LBM), which is a mesoscopic simulation approach based on the kinetic equation of the discrete-velocity distribution function [3–6]. In the limit of incompressible flows at low Knudsen number, the LBM can recover solutions of the Navier-Stokes equations [7]. The discrete lattice Boltzmann equation in velocity space with the Bhatnagar-Gross-Krook (BGK) collision operator [8] reads as

$$f_i(\mathbf{x} + \mathbf{e}_i \Delta t, t + \Delta t) - f_i(\mathbf{x}, t) = -\frac{\Delta t}{\tau}[f_i(\mathbf{x}, t) - f_i^{eq}(\mathbf{x}, t)], \tag{1}$$

where $i = 1, \ldots, 19$ for the D3Q19 model used here, $f_i(\mathbf{x}, t)$ is the particle distribution function for a single fluid component, $\mathbf{e}_i$ is the discrete velocity in the $i$th direction, $\Delta t$ is the discrete time step. $f^{eq}$ is the equilibrium function corresponding to the truncated Maxwell-Boltzmann distribution and $\tau$ is the dimensionless relaxation time which is related to the kinematic viscosity $\nu = c_s^2[\tau - \Delta t/2]$. Here $c_s = 1/\sqrt{3}\Delta x/\Delta t$ is the lattice speed of sound and $\Delta x$ is the lattice constant. The macroscopic fluid density and velocity are calculated from the moments of the distribution functions [6]. A source term can be added to the RHS of Eq. (1) to account for the contributions of the external body force and membrane forces [9, 10].

The fluid-structure interaction is handled using the immersed boundary method (IBM) [11]. This method enables the coupling of the membrane forces with the fluid flow, where forces on the membrane nodes are spread to fluid nodes and the fluid velocity is interpolated back to the membrane [12]. Membrane advection follows an explicit Euler scheme [9].

The RBC is modeled using a triangular meshed capsule with a biconcave shape as its reference undeformed state. The membrane thickness is considered to be negligible with respect to the RBC radius; thus, the strain energy of the membrane can be modeled with a 2D hyperelastic constitutive equation. We use here the Skalak law for inextensible elastic membranes [13, 14], which reads as

$$W_s = \int_A \frac{G_s}{4}[I_1^2 + 2I_1 - 2I_2 + CI_2^2]dA, \tag{2}$$

where $\int_A$ is an integral over the surface of the RBC, $I_1$ and $I_2$ are the two strain invariants, $G_s$ is the shear elastic modulus, $G_a = [1 + 2C]G_s$ is the area dilatation modulus and $C$ is a dimensionless parameter that should be set to a large value to fulfill the local inextensibility condition of the membrane. To enforce the global area ($A$) and volume ($V$) of the RBC, we introduce penalty functions controlled through the free parameter $\kappa_a$ and $\kappa_v$ and reading as

$$W_a = \frac{\kappa_a}{2}\frac{[A - A_0]^2}{A_0}, \quad W_v = \frac{\kappa_v}{2}\frac{[V - V_0]^2}{V_0}, \tag{3}$$

where $A_0$ and $V_0$ are the reference volume and area of the undeformed RBC. Forces at each membrane node ($\mathbf{r}$) are derived using virtual work principles

$$\mathbf{F}_a = -\frac{\partial W_a}{\partial \mathbf{r}}, \tag{4}$$

where the subscript $a$ refers to either $s$, $a$, or $v$ for the shear elastic force, the surface area force, or the volume force, respectively. The shear elastic force is evaluated using a linear finite element method (FEM) [12]. In addition to shear elasticity, impermeability, and inextensibility, the RBC's membrane exhibits a resistance to out-of-plane deformation described by the Helfrich free energy [15] as

$$W_b = \frac{\kappa_b}{2}\int_A [2H - c_0]^2 dA, \tag{5}$$

where $H = \frac{1}{2}(c_1 + c_2)$ and $c_0$ are the mean and spontaneous curvatures, $c_1$ and $c_2$ are the two principal curvatures, and $\kappa_b$ is the bending modulus. $c_0$ is neglected in this study. The bending force, which is obtained from the functional derivative of Eq. (5), reads as

$$\mathbf{F}_b = 2\kappa_b[2H\{H^2 - K\} + \Delta_s H]\mathbf{n}, \tag{6}$$

where $K = c_1 c_2$ is the Gaussian curvature, $\Delta_s$ is the Laplace-Beltrami operator and $\mathbf{n}$ is the outer normal vector. The discrete form of Eq. (6) is computed using discrete differential geometry operators following the approach described in [16, 17].

## 4. Numerical setup

We investigate the flow of red blood cells in a duct with a rectangular cross-section of dimensions $L_x = 60\mu m = 120\Delta x$, $L_y = 30\mu m = 60\Delta x$, and lengths $L_z = 400\mu m = 800\Delta x$ or $2400\mu m = 4800\Delta x$. The radius of the RBC is $R = 4\mu m = 8\Delta x$, and thus the lattice resolution $\Delta x = 0.5\mu m$. The RBC surface is discretized with a triangular mesh containing 2,000 faces and 1002 nodes. The tube hematocrit is varied from approximately 2% to 40%. For the largest channel length, $L_z = 4,800\Delta x$, a tube hematocrit of 40% corresponds to a total number of RBCs of around 18000, while for $L_z = 800\Delta x$, the total number of RBCs per simulation is 3000. For the shortest channel length, we perform statistics over 3 to 4 different simulations to ensure that the number of rigid RBCs is sufficient. The bare fluid has a density $\rho = 10^3 Kg/m^3$ and kinematic viscosity $\nu = 1.004 \times 10^{-6} m^2/s$. No-slip boundary conditions are applied on the $yz$ and $zx$ planes using the halfway bounce-back method [6], while periodic conditions are used along the $z$-axis. The flow is driven by a body force ($f_{b,z}^{ext}$) along the $z$-axis, resulting in a parabolic flow profile with a midplane velocity, $u_{max}$, defined as

$$u_{max} = \frac{f_{b,z}^{ext}}{2\rho\nu}\left[\frac{L_x^2}{4} - \frac{8L_x^2}{\pi^3}\sum_n^{odd}\frac{1}{n^3}\frac{\sin(n\pi/2)}{\cosh(n\pi L_y/2L_x)}\right]. \quad (7)$$

Figure 4 depicts the channel geometry used in this study and an example of the margination of the stiffened RBCs as a function of the volume fraction.

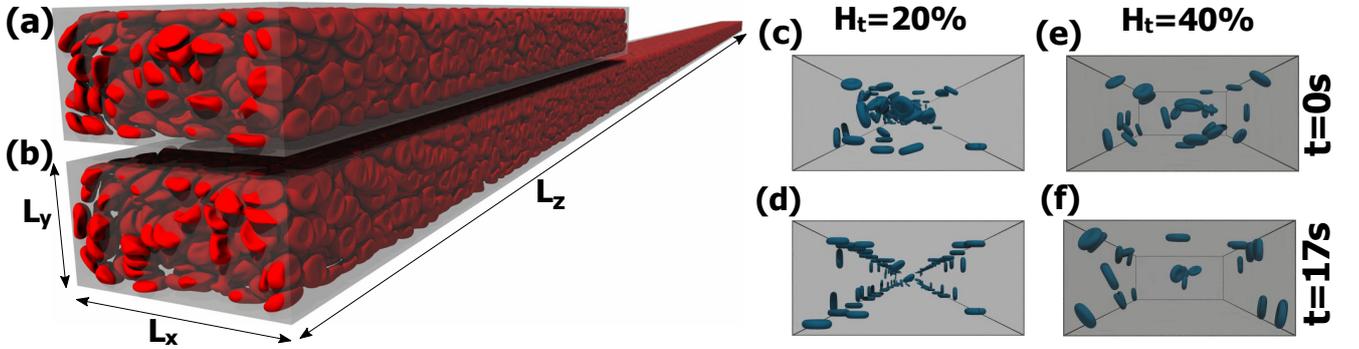

Figure 4. A suspension of red blood cells flowing through a rectangular duct channel. The width and height of the duct are $L_x = 60\mu m$ and $L_y = 30\mu m$. The length of the duct varies from $L_z = 400\mu m$ (a) to $L_z = 2400\mu m$ (b). The position of the stiffened erythrocytes (depicted in blue) subject to a parabolic flow with a bare fluid midplane velocity around 4mm/s is shown in (c) and (d) for a channel hematocrit $H_t = 20\%$ at times $t = 0$s and $t \approx 17$s. The physiological erythrocytes are not shown for clarity. (e) and (f) are similar to (c) and (d), respectively, but for a tube hematocrit $H_t = 40\%$. Note that the duct length in (c) and (d) is $L_z = 2,400\mu m$, and $L_z = 400\mu m$ in (e) and (f).

Our simulations focus on the flow of a bidisperse suspension of healthy and stiffened red blood cells (RBCs). The shear elastic and bending moduli for healthy RBCs are set to $G_s \approx 2 \cdot 10^{-6}$N/m [18, 19] and $\kappa_B \approx 2\times 10^{-19}$Nm [20], respectively. These values are 100 times higher for the stiffened RBCs, ensuring that their shape remains biconcave throughout the simulation. Stiffened RBCs comprise approximately 1% of the total cells in our simulations. The viscosity ratio between the fluid encapsulated within the RBC and the suspending fluid is fixed to $\lambda = 5$. The capillary number that describes the deformability of the RBCs is defined as $Ca \equiv \rho\nu R\dot{\gamma}_w/G_s$, with $\dot{\gamma}_w = 4u_{max}/L_x$ is the wall shear rate.

---

## II. ACKNOWLEDGMENTS

R. Chachanidze and O. Aouane as well as M. Leonetti and C. Wagner contributed equally to this work. The authors want to thank Flormann D., A.Christ and E.Terriac for their help with AFM setup and confocal microscopy. We thank T.John for the fruitful discussion. The authors gratefully acknowledge the support from the German Research Foundation (DFG FOR 2688, projects WA 1336/12-2 and HA 4382/8-1), the National Centre for Space Studies (CNES, AAP Margination) and the French National Research Agency (Polytranflow ANR-13-BS09-0015-01). R.D.C. was supported by by a fellowship from Polytransflow ANR and from the Franco-German University and a GradUS Global scholarship. The Jülich Supercomputing Centre (JSC) and the Regional Erlangen National High Performance Computing Center (NHR@FAU) are highly acknowledged for providing the required supercomputing resources.